

\input tables
\magnification=\magstep1
\def\ltwid{\mathrel{\raise.3ex\hbox{$<$\kern-.75em\lower1ex\hbox{$\sim$}}}}
\def\half{{1 \over 2}}

\rightline{UCLA/94/TEP/44}
\rightline{UFIFT-HEP-94-17}
\bigskip

\centerline {\bf CASIMIR FORCES BETWEEN BEADS ON STRINGS AND MEMBRANES}

\bigskip
\bigskip
\bigskip

\centerline {{Eric D'Hoker}
\footnote{*} {E-mail address : dhoker@uclahep.physics.ucla.edu}}
\centerline {{\it Physics Department}}
\centerline {{\it University of California}}
\centerline {{\it Los Angeles, CA 90024-1547}}
\bigskip
\medskip
\centerline {{Pierre Sikivie}
\footnote{**} {E-mail address : sikivie@ufhepa.phys.ufl.edu }}
 \centerline {{\it Physics Department}}
\centerline {{\it University of Florida}}
\centerline {{\it Gainesville, FL 32611-8440}}
\centerline {and}
\centerline {{\it Institute for Theoretical Physics}}
\centerline {{\it University of California}}
\centerline {{\it Santa Barbara, CA 93106-4030}}
\bigskip
\medskip
\centerline {{Youli Kanev}
\footnote{***} {E-mail address : kanev@phys.ufl.edu}}
\centerline {{\it Physics Department}}
\centerline {{\it University of Florida}}
\centerline {{\it Gainesville, FL 32611-8440}}

\bigskip
\bigskip
\bigskip

\centerline {\bf Abstract}

\bigskip

We develop a general formalism to calculate the force
between beads attached to a flat $d$-dimensional membrane
due to the quantum fluctuations of the membrane.
The interaction potential is derived as a function of $d$
and the membrane energy density, tension, stiffness and temperature.
We find that the induced interactions turn off when $d$ exceeds a
certain critical dimension.
The potential is attractive in all cases where it is non-zero
and at finite temperature falls off exponentially for large distances.

\vfill\eject

In a previous paper [1], two of us derived the force between
two beads on a straight string at rest due to the quantum fluctuations of the
string.  We obtained the interaction potential by explicitly summing
the zero-point energies of all the
string modes.  In the present paper we generalize our results to the case of
beads on 2-dimensional and 3-dimensional membranes,
to allow for stiffness of the string or
membrane, and to allow for finite temperature.  The methods
we use in this paper are much more efficient than those of Ref. [1].  They
yield a set of formulae which can be directly applied to a large class of
problems of this type.
Related issues have recently been discussed in the papers of ref. [2].

The physical system we consider is a straight string or flat membrane, at rest,
to which are attached masses $m_j$ at locations $\vec x_j~(j=1,2...N)$.  Let
$d$ be the dimension of the membrane and $D$ the dimension of the (flat) space
in which the membrane is embedded.  For example, for a string in physical
space,
$d=1$ and $D=3$.  The membrane has $D-d$ transverse and $d$ longitudinal
directions in which to oscillate.  For the sake of simplicity, we will first
assume that the beads can slide freely along
the membrane.  In that case, the longitudinal oscillations of the membrane
decouple from the beads and we only have to consider its transverse
oscillations when calculating the interaction potential between the beads.
We will discuss the contribution from the longitudinal oscillations, in case
the beads are stuck at a particular location on the membrane, at the end of the
paper.

The action describing the dynamics of the
displacement $\varphi (\vec x,t)$ of the
membrane in any one of its $D-d$ transverse directions is
$$
S = \int dt \int d^d x {1\over 2} \biggl [ \epsilon \biggl({\partial
\varphi\over \partial t}\biggr)^2 - \tau (\vec\nabla \varphi)^2 - \mu
(\vec\nabla\cdot\vec\nabla\varphi)^2
+ \sum_{j=1}^N m_j \biggl({\partial\varphi\over\partial t}\biggr)^2 \delta^d
(\vec x - \vec x_j)\biggr]\ ,
\eqno(1)
$$
where $\epsilon$ is the energy/volume of the membrane, $\tau$ its tension and
$\mu$ its stiffness.
For the sake of definiteness, we are neglecting at first
all higher derivative terms
beyond the one due to stiffness.  Using standard field theory methods [3], one
may derive the following formula for the energy $E$ of the lowest energy state
of the membrane with beads attached:
$$
-E = (D-d) \lim_{T_E\to\infty} {1\over 2T_E}~e^{\displaystyle{-{\sum_{j=1}^N}
{m_j\over 2} \int_{-T_E}^{T_E} dt \biggl( {\partial\over\partial t}
{\delta\over
\delta J(\vec x_j ,t)}\biggr)^2}} Z[J]\bigg\vert_{J=0\atop{\rm connected}}
\eqno(2)
$$
where
$$
Z[J] = e^{{1\over 2} \int d^{d+1} x\int d^{d+1} y J(x) \Delta_F (x-y)
J(y)}
\eqno(3)
$$
and
$$
\Delta_F (x) = \int {d^{d+1} q\over (2\pi)^{d+1}}~~ {e^{iq\cdot x}\over
\epsilon
q_0^2 + \tau \vec q\cdot \vec q + \mu (\vec q \cdot \vec q)^2}\ .
\eqno(4)
$$
The only connected graphs that appear on the RHS of Eq. (2) are single loops
with an arbitrary number of insertions.  The insertions describe the
emission/absorption of two virtual $\varphi$ quanta by any one of the $N$
beads.
After some algebra, one finds that the RHS of Eq.~(2) can be expressed
in terms of a single integral as follows:
$$
\eqalign{
-E &= (D-d) \sum_{p=1}^\infty {(-1)^p\over 2p} \sum_{j_1=1}^N ...
\sum_{j_p=1}^N m_{j_1} ... m_{j_p} \times\cr
&\times \int_{-\infty}^{+\infty} {dw\over 2\pi}~w^{2p} G(w, \vert \vec x_{j_1}
-
\vec x_{j_2}\vert) G(w,\vert \vec x_{j_2} - \vec x_{j_3} \vert) ...
G(w,\vert \vec x_{j_p} - \vec x_{j_1} \vert)\cr
&= -(D-d) \int_0^\infty {dw\over 2\pi} \ell n~det~(1+w^2 M),\cr}
\eqno(5)
$$
where $M$ is the $N \times N$ matrix of elements
$M_{jk} = m_j G(w,\vert\vec x_j -
\vec x_k \vert)$, and:
$$
G(w,\vert \vec x \vert) =G(w,|\vec x |;\epsilon,\tau,\mu)=
 \int {d^d q\over
(2\pi)^d} {e^{i\vec q\cdot\vec x}\over \epsilon w^2 + \tau \vec q\cdot\vec q +
\mu (\vec q\cdot\vec q)^2}\ .
\eqno(6)
$$
The calculation which yields Eq. (5) is similar to that which yields the
one-loop effective potential of scalar field theories [4].  In particular, the
combinatorial factor ${1\over 2p}$ for a loop with $p$ insertions appears in
both cases.  It is the product of ${1\over p!}$ from the expansion of the
exponential in Eq.~(2) and ${1\over 2} (p-1)!$ which is the number of
inequivalent cyclical orderings of $p$ objects.

The two-body interaction can be readily obtained from Eq.~(5) with N=2:
$$
\eqalign{
V (r) &= E^{N=2} (r) - E^{N=2} (\infty)\cr
&= (D-d) \int_0^\infty {dw\over 2\pi} \ell n \left[ 1- {m_1m_2w^4
G(w,r)^2\over (1+m_1w^2G(w,0))(1+m_2w^2G(w,0))} \right]\ .\cr}
\eqno(7)
$$
Eq.~(7) is far more general and simpler than the corresponding formula in
Ref.~[1].  The three-body interaction potential
$V_3(\vert\vec x_1-\vec x_2\vert, \vert\vec x_2-\vec x _3\vert, \vert
\vec x_1-\vec x_3\vert)$ can be similarly obtained from Eq.~(5) with N=3,
and so on.

For $\mu=0$ (pure tension), we have:
$$
G(w,r) = {1\over 2\sqrt{\epsilon\tau}w}~e^{-\sqrt{{\epsilon\over \tau}}
wr}~~~~ \qquad {\rm for~~d=1}
\eqno(8.1)
$$
$$
G(w,r)= {1\over 2\pi\tau} K_0\left(\sqrt{{\epsilon\over\tau}}
wr\right)~~ \qquad {\rm for~~d=2}
\eqno(8.2)
$$
$$
G(w,r)= {1\over 4\pi\tau r} e^{-\sqrt{{\epsilon\over\tau}}
wr}~~~~~~~~~ \qquad {\rm for~~d=3}
\eqno(8.3)
$$
where $K_0(x)$ is the modified Bessel function of the third kind.  The
$G(w,r)$
for $\mu \not= 0$ can be obtained from those for $\mu=0$ using:
$$
G(w,r;\epsilon,\tau,\mu) = {1\over \mu (q_2^2 - q_1^2)} [G(q_1,r;1,1,0) -
G(q_2,r;1,1,0)]
\eqno(9)
$$
where
$$
q_{1\atop 2}^2 = {1\over 2\mu} \biggl[ \tau \mp
\sqrt {\tau^2 -4\epsilon\mu w^2}\biggr]
\eqno(10)
$$
and $Re~q_1,~Re~q_2>0$.  For the case of pure stiffness $(\tau =0)$, one has:
$$
G(w,r) = {e^{-\sqrt{{w\over 2}\sqrt{{\epsilon\over \mu}}} r}\over
2\sqrt {2w^3 \sqrt{\mu\epsilon^3}}} \biggl[ cos\biggl(\sqrt{{w\over 2}
\sqrt{{\epsilon\over\mu}}} r\biggr) + sin \biggl( \sqrt{{w\over 2}
\sqrt{{\epsilon\over\mu}}} r\biggr)\biggr]~~~~~~~~~~~~~~{\rm
for~d=1}
\eqno(11.1)
$$
$$
G(w,r)={1\over 4\pi i w \sqrt{\epsilon\mu}}\biggl[ K_0 \biggl( (1-i)
\sqrt{{w\over 2}\sqrt{{\epsilon\over\mu}}} r\biggr) -K_0 \biggl( (1+i)
\sqrt{{w\over 2}\sqrt{{\epsilon\over\mu}}}
r\biggr)\biggr]~~~~~~~{\rm for~d=2}
\eqno(11.2)
$$
$$
G(w,r)={e^{-\sqrt{{w\over 2}\sqrt{{\epsilon\over \mu}}} r}\over 4\pi r w
\sqrt{\epsilon\mu}}~sin \biggl( \sqrt{{w\over 2}\sqrt{{\epsilon\over\mu}}}
r\biggr)~~~~~~~~~~~~~~~~~~~~~~~~~~~~~~~~~~~~~~~~~~~~~{\rm for~~d=3.}
\eqno(11.3)
$$
Note that $G(w,0)$ is finite in all cases except
when $\mu=0$ and $d=2,3$.
In these two cases $G(w,r) \sim (\ell n~r),~r^{-1}$
respectively as $r\to 0$.
It can be seen from Eq.~(7) that $V (r) = 0$ when $G(w,0) = \infty$.
In the two cases where $G(w,0) = \infty$, we introduce a
short-distance cut-off $\delta$.
In many physical applications such a cut-off is in fact present
because of the thickness of the membrane.
In particular, a small stiffness acts as a cutoff
with $\delta = \sqrt {\mu /\tau }$.
The role of the cut-off is to restrict the region of
integration in Eq.~(6) to $q \ltwid \delta^{-1}$.
This in effect replaces $G(w,0)$ by $G(w,\delta)$.

Using Eqs.~(7-11), we have calculated the interaction potential $V_2(r)$ in a
variety of cases.  The results are given in Table I.  The numerical
coefficients
$C_1...C_3$ that appear there are defined by:
$$
C_1 \equiv - \int_0^\infty dy~y~\ell n [1-e^{-y} (1+sin~y)] = 2.34
\eqno(12.1)
$$
$$
C_2 \equiv - 4\int_0^\infty dy~y~\ell n \biggl[ 1-\biggl({2\over i\pi}\bigl[
K_0\bigl( (1-i)y\bigr) - K_0 \bigl(
(1+i)y\bigr)\bigr]\biggr)^2\biggr]=1.84
\eqno(12.2)
$$
$$
C_3 \equiv - \int_0^\infty dy~y~\ell n \biggl[ 1 - {4\over y^2} e^{-y} sin^2
\bigl({y\over 2}\bigr)\biggr] =0.86\ .
\eqno(12.3)
$$
\bigskip

\begintable
{}~~~~~~~|d=1|d=2|d=3\cr
$\displaystyle{\mu=0\atop {\rm small~m}}$|
$\displaystyle{{-m_1m_2\over 32\pi\epsilon^2 r^3}
({\tau\over\epsilon})^{1/2}}$|
$\displaystyle{{-27m_1m_2\over
2^{12}\pi\epsilon^2r^5}({\tau\over\epsilon})^{1/2}}$|
$\displaystyle{{-3m_1m_2\over2^7\pi^3\epsilon^2 r^7}
({\tau\over\epsilon})^{1/2}}$
\cr

$\displaystyle{\tau=0\atop {\rm small~m}}$|
$\displaystyle{{-3m_1m_2\over 16\pi\epsilon^2 r^4}
({\mu\over\epsilon})^{1/2}}$|$\displaystyle{{- m_1m_2\over 4\pi^2\epsilon^2
r^6}
({\mu\over\epsilon})^{1/2}}$|$\displaystyle{{-15 m_1m_2\over 32\pi^3\epsilon^2
r^8}
({\mu\over\epsilon})^{1/2}}$\cr

$\displaystyle{\mu=0\atop {\rm large~m}}$|$\displaystyle{{-\pi\over 24r}
({\tau\over\epsilon})^{1/2}}$|$\displaystyle{{-\pi\over 8r} {1\over (\ell
n\delta)^2}
({\tau\over\epsilon})^{1/2}}$|$\displaystyle{{-\delta^2\over 4\pi r^3}
({\tau\over\epsilon})^{1/2}}$\cr

$\displaystyle{\tau=0\atop {\rm large~m}}$|$\displaystyle{{-C_1\over 2\pi r^2}
({\mu\over\epsilon})^{1/2}}$|$\displaystyle{{-C_2\over 2\pi r^2}
({\mu\over\epsilon})^{1/2}}$|$\displaystyle{{-C_3\over 2\pi r^2}
({\mu\over\epsilon})^{1/2}}$\endtable

\medskip

\noindent {\sl Table I:} The contribution to the two-bead interaction
potential due to the transverse oscillations of the membrane, in a
variety of cases.  To obtain $V^\perp (r)$, multiply each entry by
$\hbar (D-d)$.  To obtain the contribution $V^{\vert\vert} (r)$ due
to longitudinal oscillations, replace $\epsilon$ by
$\epsilon - \tau$ and $({\tau\over\epsilon})^{1/2}$ by $(-{d\tau\over
d\epsilon })^{1/2}$, and multiply each entry by $\hbar d$.

\vskip .25truein

The cases labeled ``small $m$'' in Table I are when
$m_i~w^2 G(w,0)<< 1~ (i=1,2)$ inside the
integral of Eq.~(7) and the interaction potential can be approximated by
$$
V (r) = -(D-d) {m_1m_2\over 2\pi} \int_0^\infty dw~w^4 G(w,r)^2\ .
\eqno(13)
$$
The cases labeled ``large $m$'' in Table I  is when $m_i w^2 G(w,\delta )
\gg 1$ $(i=1,~2)$ inside the integral of Eq. (7) and the interaction
potential can be approximated by
$$
V (r) = (D-d)\int_0^\infty {dw\over 2\pi} \ell n \biggl[
 1-\bigl(
{G(w,r)\over G(w,0)}\bigr)^2\biggr]\ .
\eqno(14)
$$
The critical
parameter which distinguishes the small $m$ and large $m$ regimes is ${m\over
\epsilon r^d}$ in all cases where $G(w,0)$ is finite.  For the two cases where
$G(w,0)$ is infinite, and a short-distance cut-off $\delta$ was introduced
as discussed above, the critical parameter is: ${m\over \epsilon r^2}
\ell n ({r\over \delta})$ for $\mu=0,~d=2$ and ${m\over \epsilon r^2\delta}$
for $\mu=0,~d=3$.  Note that in these two cases one necessarily enters the
large
mass regime when $\delta\to 0$ and that $V (r)$ vanishes
in that limit as $(\ell n\delta)^{-2}$ and $\delta^2$ respectively.

We notice that whenever the potential is non-zero, it is attractive for
all values of $r$ and produces an instability under which beads tend
to group together in lumps. This type of instability was already noticed
for the
special case of strings ($d=1$) in [1].

All the results of Ref.~[1] except one (see below) are easily reproduced using
Eqs.~(7) and (8.1) and the generalization of Eq.~(7) for N=3.  One statement in
Ref.~[1] cannot be reproduced because it is in error.  It concerns the
next-to-leading term in $V (r)$ for the case of large masses and d=1,
$\mu=0$.  The correct answer is:
$$
V (r) = (D-1) ({\tau\over\epsilon})^{1/2} \biggl[ -{\pi\over 24r}
+ {1\over 2} \sqrt {{(m_1+m_2)\epsilon\over m_1m_2 r}} + ...\biggr]\ .
\eqno(15)
$$
Higher order corrections of various sorts may be derived from Eq.~(7).  For
example the expansion of $V (r)$ in small masses to 3rd order for d=1,
$\mu=0$ is:
$$
\eqalign{V (r) = &- (D-1) {1\over 32\pi} ({\tau\over \epsilon})^{1/2}
{m_1m_2\over \epsilon^2 r^3} \biggl\{ 1- {3\over 4} {m_1+m_2\over \epsilon
r}\cr
&+\bigl[ {3\over 4} (m_1^2 + m_2^2) + {195\over 256} m_1 m_2\bigr] {1\over
\epsilon^2 r^2} + ...\biggr\}\ .\cr}
\eqno(16)
$$
The lowest order correction to the interaction potential in the case of pure
tension, d=1 and small masses, due to a small but non-zero $\mu$ is:
$$
V (r) = - (D-1) \sqrt{{\tau\over\epsilon}} {m_1m_2\over 32\pi
\epsilon^2 r^3} (1+ {3\mu\over 2r^2} + ...)\ .
\eqno(17)
$$
Et cetera.

At non-zero temperature $T$, the interaction potential equals the free energy
which is given by
$$
F(T) = - T (D-d) e^{-\sum_{j=1}^N {m_j\over 2} \int_0^{1/T} dt ({\partial\over
\partial t} {\delta\over \delta J(\vec x_j,t)})^2} W[J]\bigg\vert_{J=0\atop
{\rm connected}}\ .
\eqno(18)
$$
In the expression for $W[J]$, Eq. (3), $\Delta_F(x)$ must now be replaced by:
$$
\Delta_T (x) = \sum_{n=-\infty}^{+\infty} T\int_{-\infty}^{+\infty}
{d^dq\over (2\pi)^d} {e^{i(\vec q\cdot \vec x + 2\pi nTt)}\over
\epsilon(2\pi nT)^2 + \tau\vec q\cdot\vec q + \mu (\vec q\cdot\vec
q)^2}\ .
\eqno(19)
$$
Eq.~(5) becomes:
$$
F = (D-d) T \sum_{n=1}^\infty \ell n~det (1+w_n^2 M)
\eqno(20)
$$
with $w_n = 2\pi nT$ and $M_{jk} = m_j G(w_n,\vert\vec x_j -\vec
x_k\vert)$.  Thus, to allow for finite $T$, one simply replaces the integral
over $w$ by a sum over the discrete frequencies $w_n = 2\pi nT$.  In
particular:
$$
V (r) = (D-d) T \sum_{n=1}^\infty \ell n \biggl[ 1-
{m_1m_2w_n^4 G(w_n,r)^2\over
\bigl(1+m_1w_n^2G(w_n,0)\bigr)\bigl(1+m_2w_n^2 G(w_n,0)\bigr)}\biggr]\
.\eqno(21)
$$
For finite $T$ and large $r$, $G(w_n,r)$ falls off exponentially, as $e^{-2\pi
nrT\sqrt{{\epsilon\over\tau}}}$ if $\mu=0$ and as $e^{-\sqrt{\pi
nT\sqrt{{\epsilon\over\mu}}} r}$ if $\tau=0$.  In that limit the $n=1$
term dominates over all others and
$$
V (r) = -(D-d) T {m_1m_2 (2\pi T)^4 G(2\pi T,r)^2\over
\bigl(1+2\pi m_1 T G(2\pi T{,}0)\bigr)\bigl(1+2\pi m_2 T G(2\pi T{,}0)\bigr)}\
{}.
\eqno(22)
$$
One needs only to substitute the expressions for $G(2\pi T,r)$,
Eqs. (8-11), to obtain the various interaction potentials in that limit.
In all cases, the
interactions between beads on strings and membranes becomes short ranged
as soon as the temperature is non-zero.
For small masses and all values of $T$,
$$
V (r) = -(D-d) T m_1m_2 \sum_{n=1}^\infty w_n^4 G(w_n,r)^2\
.\eqno(23)
$$
In some cases, the sum can be done explicitly; e.g. for $d=1$ and $\mu
=0$:
$$
V (r) = -(D-1) {\pi^2 m_1m_2 T^3\over 4\epsilon \tau} {cosh
(2\pi Tr\sqrt{{\epsilon\over\tau}})\over sinh^3 (2\pi Tr
\sqrt{{\epsilon\over\tau}})}\ .
\eqno(24)
$$
For large masses and all values of $T$,
$$
V (r) =  (D-d) T \sum_{n=1}^\infty \ell n \biggl[ 1-\biggl(
{G(w_n,r)\over G(w_n,0)}\biggr)^2\biggr]\ .
\eqno(25)
$$
In the case $d=1$ and $\mu =0$, this yields
$$
V (r) =  (D-1) T ~\ell n\prod_{n=1}^\infty (1-\alpha^n)
\eqno(26)
$$
where $\alpha = e^{-4\pi Tr\sqrt{{\epsilon\over\tau}}}$.
This expression is well-known from string theory and dual models
and exhibits a duality symmetry under the interchange of $r$ and
$1/r$.

Let us turn to the contribution from longitudinal oscillations
of the membrane.  Such a contribution exists only if the beads are stuck
to the membrane, i.e. they are constrained not only to stay on the
membrane but to stay at a particular location on the membrane.  In
particular, the contribution vanishes in the limit $\tau = \epsilon$,
where the membrane becomes invariant under boosts parallel to its
surface.  The relevant action is:
$$
S= \int dt \int d^d x \bigg\{ {\epsilon -\tau\over 2}
\biggl[ ({\partial\varphi\over \partial t})^2 - v_L^2 (\vec\nabla
\varphi)^2 \biggr ] -\mu (\vec \nabla \cdot \vec \nabla \varphi )^2
+\sum_{j=1}^N {m_j\over 2} ({\partial\varphi\over\partial t})^2
\delta^d (\vec x-\vec x_j)\biggr\}
\eqno(27)
$$
where $\varphi$ is the displacement of the membrane from its equilibrium
position in any one of the $d$ longitudinal directions and $v_L^2 = -
{d\tau\over d\epsilon}$ is the (velocity)$^2$ of longitudinal
oscillations.  Here we are neglecting all terms with more than
two spatial derivatives.  The action of Eq.~(27) reproduces the
equations of motion and stress-energy-momentum tensor of the
longitudinal oscillations [5].  Comparison of Eqs.~(1) and (27) shows that one
can obtain the contribution due to longitudinal oscillations from that due to
transverse oscillations by replacing $\epsilon\to\epsilon -\tau,~\tau\to
-(\epsilon -\tau) {d\tau\over d\epsilon},~D-d\to d$.

\bigskip

Finally, let us point out a generalization in which the oscillations of
the $d$-dimensional membrane have a general dispersion
law : $\epsilon w ^2 = P(\vec q \cdot \vec q)$, where $P$ is any positive
function of $\vec q \cdot \vec q$ which vanishes when $\vec q \to 0$.
When only tension and stiffness terms were present in the membrane action,
$P$ is a polynomial of second degree in $\vec q \cdot \vec q$
with vanishing constant term. However, one may imagine many other possible
dispersion laws, including arbitrarily high powers of $\vec q \cdot \vec q$
or even fractional powers of $\vec q \cdot \vec q$. This
may occur when the membrane Lagrangian is viewed as an effective
theory resulting from complicated microscopic dynamics.
The formalism developed above may be applied to this generalized
case in a straightforward way : formulas (5) and (7) are unchanged
and in the propagators of (4) and (6), it suffices to replace
$\tau \vec q \cdot \vec q + \mu (\vec q \cdot \vec q)^2$ by
$P(\vec q \cdot \vec q)$.
In particular, we have the Green function
$$
G(w,\vert \vec x \vert) =
\int {d^d q\over (2\pi)^d}
{e^{i\vec q\cdot\vec x}\over \epsilon w^2 + P(\vec q~^2)}\ .
\eqno(28)
$$

We shall examine the behavior of the two-body potential when $P$
approaches the following limits
$$
\eqalign{
P(\vec q \cdot \vec q) &\sim P_\infty \times (\vec q \cdot \vec q)
^{\alpha _\infty }
\qquad {\rm for}
\qquad \vec q  \to \infty \cr
&\sim P_0 \times (\vec q \cdot \vec q) ^{\alpha _0 } ~~\qquad {\rm for}
\qquad \vec q  \to 0 \cr}
\eqno (29)
$$
When $2\alpha _\infty \leq d$, we have $G(w,0)=\infty$ and as a result the
entire two-body potential vanishes identically.
We note therefore that there is a critical dimension above which the
fluctuating membrane is unaffected by the extra masses.
In particular, as was already noted, the potential vanishes for $d=2,~3$
and higher in the case of pure tension.
Furthermore, the potential vanishes for $d \geq 4$, if stiffness is the highest
derivative term in the membrane action.
The presence of a cutoff
produces finite but non universal interactions.

If $2 \alpha _\infty > d$, $G(w,0)$ is finite and we may use
the formalism developed above to determine the long and short
distance behaviors of the interaction potential corresponding to the
limiting behaviors of the dispersion law given in Eq. (29).
We find
$$
\eqalign{
V(r) &= - {m_1 m_2 \over 2 \pi \epsilon ^2} \biggl ({P_0 \over \epsilon}
\biggr )^\half  {1 \over r^{2d +\alpha _0}}
\int _0 ^\infty du ~u^{2d /\alpha _0} C(\alpha _0,u)^2 , \quad
 \qquad {\rm for} \qquad
r\to \infty \cr
V(r) &= - {1 \over 2 \pi} \biggl ( { P_\infty \over \epsilon} \biggr )^\half
{1\over r^{\alpha _\infty}}
\int _0 ^\infty du ~\ell n \biggl [ 1 - {C(\alpha _\infty, u)^2
\over C(\alpha _\infty , 0)^2} \biggr ] , ~~\qquad {\rm for}
\qquad r \to 0 \cr}
\eqno (30)
$$
The function $C(\alpha , u)$ is given by
$$
C(\alpha , u) =
\int {d^d p \over (2\pi)^d} {e^{i \vec p \cdot \hat n u^{1/\alpha }}
\over 1 +(\vec p \cdot \vec p) ^{\alpha}}
\eqno (31)
$$
In Eq. (31), the function $C(\alpha , u)$ is independent of the arbitrary
unit vector $\hat n$.
This behavior may be verified on the cases that we have already presented
in Table I, and are found to be in agreement.

\bigskip
\bigskip

\noindent {\bf Acknowledgements}

\medskip

One of us (P.S.) would like to thank Larus Thorlacius and Charles B.
Thorn for stimulating discussions.  This research was supported in part by
NSF grants PHY-92-18990, PHY-89-04035 and by DOE contract No.
DE-FG05-86ER40272.

\bigskip
\bigskip

\noindent {\bf References}

\item {1.} E. D'Hoker and P. Sikivie, Phys. Rev. Lett. {\bf 71} (1993)
1136.
\item {2.} D. F\"orster, Phys. Lett. {\bf 114 A} (1986) 115;
	F. David and S. Leibler, Phys. II France {\bf 1} (1991) 959;
	M. Goulian, R. Bruinsma and P. Pincus, Europhysics Lett. {\bf
	22} (1993) 145;
	W. Cai, T.C. Lubensky, P. Nelson and T. Powers, ``Measure Factors,
	Tension and Correlations of Fluid Membranes'', UPR-599T preprint
	(1994)
\item {3.} See for example: P. Ramond, ``Field Theory: A Modern
Primer'', Addison-Wesley, 1989.
\item {4.} S. Coleman and E. Weinberg, Phys. Rev. {\bf D7} (1973) 1888.
\item {5.} See for example : J. Hong, J. Kim, and P. Sikivie,
Phys. Rev. Lett. {\bf 69} (1992) 12611; J. Kim and P. Sikivie, ``Wiggly
relativistic strings'', University of Florida preprint UFIFT-HEP-94-4
(May 1994), to appear in Phys. Rev. D.; B. Carter, ``Transonic Elastic
Model for Wiggly Goto-Nambu String", hep-th-9411231, CNRS preprint (1994)

\end